\documentclass[twocolumn,showpacs,preprintnumbers,amsmath,amssymb]{revtex4}

\usepackage{graphicx}
\usepackage{dcolumn}
\usepackage{bm}
\usepackage{color}

\begin{document}

\preprint{}

\title{A reappraisal of two-loop contributions to the fermion electric dipole moments \\
in R-parity violating supersymmetric models}

\author{Nodoka Yamanaka, Toru Sato and Takahiro Kubota}
\affiliation{%
Department of Physics, Osaka University, Toyonaka, Osaka 560-0043, Japan}%

\date{\today}

\begin{abstract}
We reexamine the R-parity violating contribution to the fermion electric and 
chromo-electric dipole moments (EDM and cEDM) in the two-loop diagrams.
It is found that the leading Barr-Zee type two-loop contribution is smaller than the result found in previous works, and that EDM experimental data provide looser limits on RPV couplings.

\end{abstract}

\maketitle

The supersymmetry is known to resolve many theoretical problems which have been encountered in the standard model (SM) such as the cancellation of the power divergences in radiative corrections, 
and its supersymmetric extension is therefore one of the promising candidates of new physics.
The supersymmetric SM can be extended to allow baryon number or lepton number violating 
interactions, known as the R-parity violating (RPV) interactions, 
and they have been constrained from the analysis of various phenomena \cite{rpvphenomenology}. 

The electric dipole moment (EDM)  is  an excellent observable 
to investigate the underlying mechanisms of the P and CP violations  
and can be measured in a variety of systems \cite{edmexp}.
Since the contribution of the SM to the EDM is in general small \cite{smedm},
it is a very good experimental observable
to examine  the supersymmetric models and other candidates of new physics.
In the past three decades, many analyses of the supersymmetric models 
with \cite{susyedm1-loop,susyedm2-loop,susyedmgeneral,susyedmflavorchange} and without 
\cite{barbieri,godbole,herczege-n,chang,choi,faessler} the conservation 
of R-parity have been done using the EDMs.

In the RPV supersymmetric model with trilinear RPV interactions, it has been
found that the fermion 
(quark or lepton) EDM does not receive any one-loop
contribution \cite{godbole}, and 
the two-loop contribution has been
analyzed in detail to give the Barr-Zee type diagram as the leading contribution \cite{chang}.
In this paper, we reexamine the RPV Barr-Zee type contribution 
which turns out to be in disagreement with previous works \cite{chang,herczege-n,faessler}.
We will show that the RPV Barr-Zee type diagram has actually a smaller contribution than that given in previous works.

The RPV interactions are generated by the following superpotential:
\begin{equation}
W_{{\rm R}\hspace{-.5em}/} =
\frac{1}{2} \lambda_{ijk} \epsilon_{ab} L_i^a L_j^b (E^c)_k
+\lambda'_{ijk} \epsilon_{ab} L_i^a Q_j^b ( D^c)_k \ ,
\label{eq:superpotential}
\end{equation}
with $i,j,k=1,2,3$ indicating the generation, $a,b=1,2$ the $SU(2)_L$ indices. 
$L$ and $E^c$ denote the lepton doublet and singlet left-chiral superfields. $Q$, $U^c$ and $D^c$ denote respectively the quark doublet, up quark singlet and down quark singlet left-chiral superfields.
The RPV baryon number violating interactions are irrelevant in this analysis since they do not contribute to the Barr-Zee type diagrams,
and are not included in our current analysis.
Also the bilinear RPV interactions were not considered.
The RPV lagrangian of interest is then given as
\begin{widetext}
\begin{eqnarray}
{\cal L }_{\rm R\hspace{-.5em}/\,} &=&
- \frac{1}{2} \lambda_{ijk} \left[
\tilde \nu_i \bar e_k P_L e_j +\tilde e_{Lj} \bar e_k P_L \nu_i + \tilde e_{Rk}^\dagger \bar \nu_i^c P_L e_j -(i \leftrightarrow j ) \right] + ({\rm h.c.})\nonumber\\
&&-\lambda'_{ijk} \left[
\tilde \nu_i \bar d_k P_L d_j + \tilde d_{Lj} \bar d_k P_L \nu_i +\tilde d_{Rk}^\dagger \bar \nu_i^c P_L d_j -\tilde e_{Li} \bar d_k P_L u_j - \tilde u_{Lj} \bar d_k P_L e_i - \tilde d_{Rk}^\dagger \bar e_i^c P_L u_j \right]  + ({\rm h.c.})  \ ,
\label{eq:rpvlagrangian}
\end{eqnarray}
\end{widetext}
where $P_L \equiv \frac{1}{2} (1-\gamma_5)$ and we also define $P_R \equiv \frac{1}{2} (1+\gamma_5)$
for later use. These RPV interactions are lepton number violating Yukawa interactions.

\begin{figure}[htbp]
\begin{center}

\begin{tabular}{cc}

\hspace{-0.5em}

\begin{minipage}{0.48\hsize}
\begin{center}
\includegraphics[width=9cm]{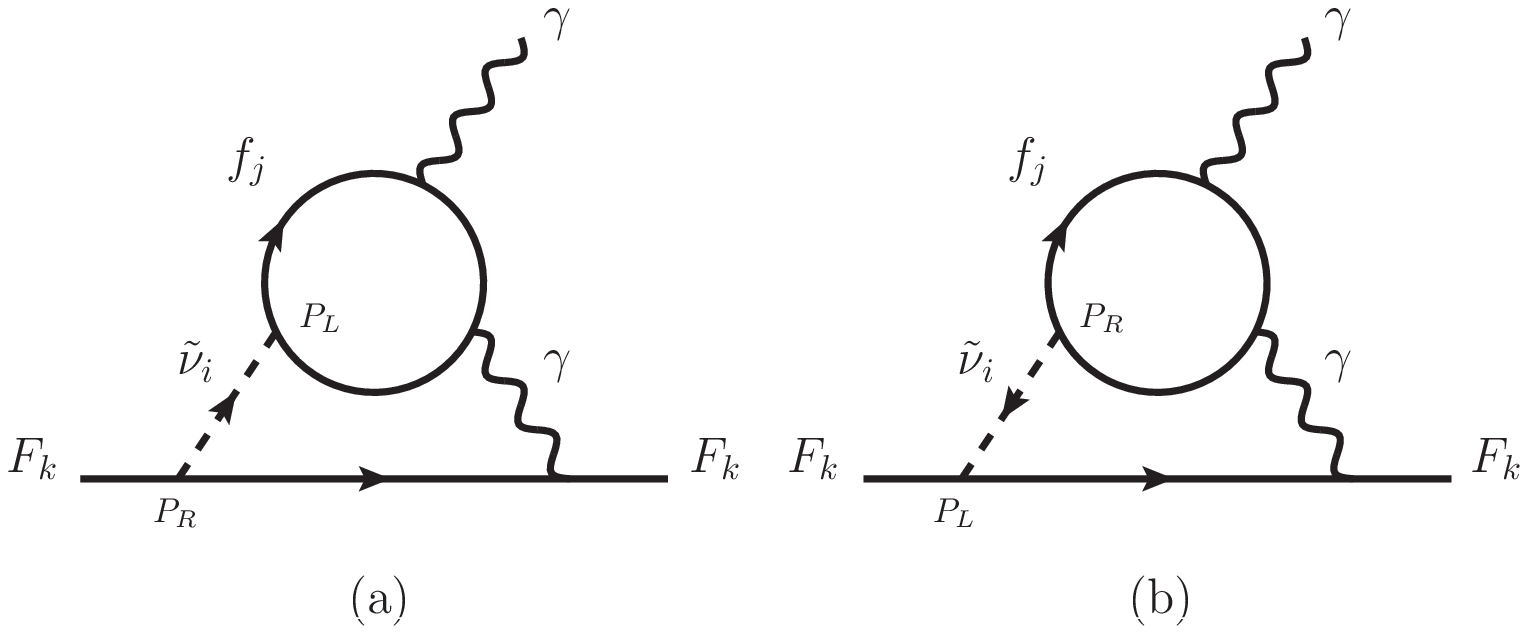}
\caption{\label{fig:barr-zee} 
Examples of Barr-Zee type two-loop contributions to the fermion EDM within RPV interactions.
The projections of the chirality ($P_L$ and $P_R$) were explicitly given for the RPV vertex.
}
\end{center}
\end{minipage}

\hspace{1.3em}

\begin{minipage}{0.48\hsize}
\begin{center}
\includegraphics[width=5cm]{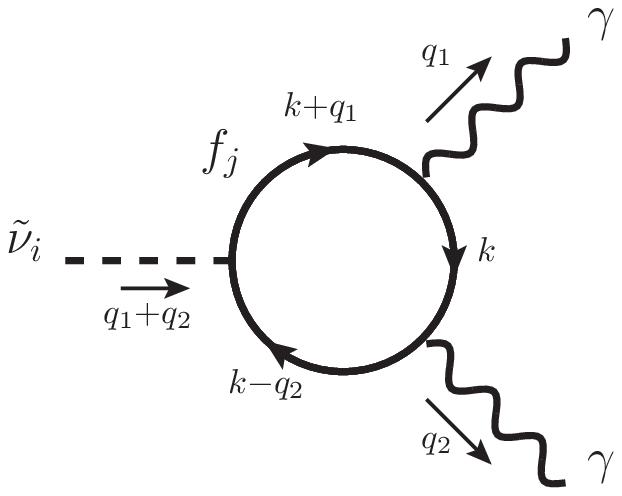}
\caption{\label{fig:nugg}
One-loop $\tilde \nu \gamma \gamma$ vertex generated with RPV interactions.
}
\end{center}
\end{minipage}
\end{tabular}

\end{center}

\end{figure}

The EDM $d_F$ of the fermion is defined as follows:
\begin{equation}
{\cal L}_{\rm EDM} = -i  \frac{d_F}{2} \bar \psi \gamma_5 \sigma^{\mu \nu} \psi F_{\mu \nu } \, ,
\end{equation}
where $F_{\mu\nu}$ is the electromagnetic field strength.
With the RPV lagrangian (\ref{eq:rpvlagrangian}), the sneutrino exchange Barr-Zee type diagrams
shown in Fig. \ref{fig:barr-zee} contribute to the EDM.
Here the emission (absorption) of the sneutrino from fermion is accompanied by $P_R$ ($P_L$) projection operator as is apparent from the Eq. (\ref{eq:rpvlagrangian}) .

%

\newpage

\onecolumngrid

\twocolumngrid

At first we give the expression of the two-photon decay amplitude of
annihilation and production of sneutrino
with internal fermion loop  shown in Fig. \ref{fig:nugg} given as
\begin{widetext}
\begin{eqnarray}
\epsilon^*_\mu (q_1) \epsilon^*_\nu (q_2) i{\cal M}^{\mu\nu}_L (q_1 , q_2) &=& 
-\hat \lambda_{ijj} n_c (Q_f e)^2 \epsilon^*_\mu (q_1) \epsilon^*_\nu (q_2)
\int \frac{d^4 k }{(2\pi)^4} \frac{{\rm Tr} \left[ (k\hspace{-.5em} / + q_1\hspace{-.9em} / \,  +m_{f_j}  ) \gamma^\mu (k\hspace{-.5em} /  +m_{f_j}  ) \gamma^\nu (k\hspace{-.5em} / - q_2\hspace{-.9em} / \,  +m_{f_j}  ) P_L \right]}{\left[(k+q_1)^2-m_{f_j}^2 \right] \left[k^2-m_{f_j}^2\right] \left[(k-q_2)^2-m_{f_j}^2 \right]} \nonumber\\
&\approx& \frac{i m_{f_j} \hat \lambda_{ijj} n_c (Q_f e)^2}{(4\pi)^2} \epsilon^*_\mu (q_1) \epsilon^*_\nu (q_2) 
\int^1_0 dx \frac{\left(1-2x(1-x) \right) \left( q_2^\mu q_1^\nu -(q_1\cdot q_2 ) g^{\mu \nu}  \right)-i\epsilon^{\mu \nu}_{ \ \ \alpha \beta} q_1^\alpha q_2^\beta}{m_{f_j}^2-x(1-x)q_1^2 } \, , \nonumber\\
\label{eq:4}
&&\\
\epsilon^*_\mu (q_1) \epsilon^*_\nu (q_2) i{\cal M}^{\mu\nu}_R (q_1 , q_2) &=& 
-\hat \lambda^*_{ijj} n_c (Q_f e)^2 \epsilon^*_\mu (q_1) \epsilon^*_\nu (q_2)
\int \frac{d^4 k }{(2\pi)^4} \frac{{\rm Tr} \left[ (k\hspace{-.5em} / + q_1\hspace{-.9em} / \,  +m_{f_j}  ) \gamma^\mu (k\hspace{-.5em} /  +m_{f_j}  ) \gamma^\nu (k\hspace{-.5em} / - q_2\hspace{-.9em} / \,  +m_{f_j}  ) P_R \right]}{\left[(k+q_1)^2-m_{f_j}^2 \right] \left[k^2-m_{f_j}^2\right] \left[(k-q_2)^2-m_{f_j}^2 \right]} \nonumber\\
&\approx& \frac{i m_{f_j} \hat \lambda^*_{ijj} n_c (Q_f e)^2}{(4\pi)^2} \epsilon^*_\mu (q_1) \epsilon^*_\nu (q_2) 
\int^1_0 dx \frac{\left(1-2x(1-x) \right) \left( q_2^\mu q_1^\nu -(q_1\cdot q_2 ) g^{\mu \nu}  \right)+i\epsilon^{\mu \nu}_{ \ \ \alpha \beta} q_1^\alpha q_2^\beta}{m_{f_j}^2-x(1-x)q_1^2 } \, , \nonumber\\
\label{eq:5}
\end{eqnarray}
\end{widetext}
where $i$ and $j$ denote the flavor indices of $\tilde \nu$ and loop fermion, respectively.
$\hat \lambda $ is the R-parity violating coupling, $\hat \lambda = \lambda $ when charged lepton runs in the loop, and $\hat \lambda = \lambda'$ in the case of down type quark. 
$n_c = 1$ ($n_c = 3$ ) if $f_j$ is a lepton (quark).
$m_{f_j}$ and $Q_f$ are the mass and the charge in unit of $e$ of the loop fermion, respectively. 
The second line is the approximated expression taking $q_2$ to be small.
To be precise, the Levi-Civita tensor is defined by 
$\epsilon^{0123} \equiv +1$, and $\gamma_5 \equiv i\gamma^0 \gamma^1 \gamma^2 \gamma^3$.

We now insert the effective $\tilde \nu \gamma \gamma$ vertices (\ref{eq:4}) and (\ref{eq:5}) into the whole Barr-Zee type diagram.
Then we end up with
\begin{widetext}
\begin{eqnarray}
i{\cal M}_{\rm BZ}
&=&
- \tilde \lambda^*_{ikk} Q_F e \, \epsilon^*_\nu (q) \int \frac{d^4 k }{(2\pi )^4} \frac{\bar u( p-q ) \gamma_\mu (p\hspace{-.5em} / - q\hspace{-.5em} / - k\hspace{-.5em}/ +m_{F_k}) P_R u(p) \cdot {\cal M}_L^{\mu \nu} (k,q)}{k^2 \left[ (q+k)^2 - m_{\tilde \nu_i }^2 \right] \left[ (p-q-k)^2-m_{F_k}^2 \right]} \nonumber\\
&&
- \tilde \lambda_{ikk} Q_F e \, \epsilon^*_\nu (q) \int \frac{d^4 k }{(2\pi )^4} \frac{\bar u( p-q ) \gamma_\mu (p\hspace{-.5em} / - q\hspace{-.5em} / - k\hspace{-.5em}/ +m_{F_k}) P_L u(p) \cdot {\cal M}_R^{\mu \nu} (k,q)}{k^2 \left[ (q+k)^2 - m_{\tilde \nu_i }^2 \right] \left[ (p-q-k)^2-m_{F_k}^2 \right]} \nonumber\\
&\approx&
i {\rm Im} ( \hat \lambda_{ijj} \tilde \lambda^*_{ikk} ) \frac{ \alpha_{\rm em} }{(4\pi )^3} n_c Q_f^2 Q_F e \frac{1}{m_{f_j}} \left\{ f \left( \frac{m_{f_j}^2}{ m_{\tilde \nu_i }^2 } \right) - g\left( \frac{m_{f_j}^2}{ m_{\tilde \nu_i }^2 } \right) \right\}  \epsilon^*_\nu (q) \bar u  \sigma^{\mu \nu} q_\mu  \gamma_5 u \, ,
\label{eq:secloop}
\end{eqnarray}
\end{widetext}
where $\tilde \lambda = \lambda$ for lepton EDM contribution and $\tilde \lambda = \lambda'$ for quark EDM contribution, and $f$ and $g$ are defined as
\clearpage
\begin{eqnarray}
f(z) &=& \frac{z}{2 } \int^1_0 dx \frac{1-2x(1-x)}{ x(1-x)-z} \ln \left( \frac{ x(1-x)}{z} \right) \, , \\
g(z) &=& \frac{z}{2 } \int^1_0 dx \frac{1}{ x(1-x)-z} \ln \left( \frac{ x(1-x)}{z} \right)  \, ,
\end{eqnarray}
in the notation of the original notation of Barr and Zee \cite{barr-zee}.
For small $z$, we have 
$g(z) \approx \frac{z}{2} 
\left( \frac{\pi^2}{3} + (\ln z)^2 \right)$, 
and 
$f(z) \approx \frac{z}{2} 
\left( \frac{\pi^2}{3}+4 + 2\ln z + (\ln z)^2 \right) $.
In the last line of Eq. (\ref{eq:secloop}), we have taken only the part 
of $i{\cal M}_{\rm BZ}$ which contributes to the EDM, disregarding 
Re($\hat \lambda_{ijj} \tilde \lambda^*_{ikk}$).
For each diagram of Fig. 1, there are also diagrams with 
the internal fermion loop reversed, and those with internal photon and 
sneutrino lines interchanged.
They all give the same amplitude and Eq. (\ref{eq:secloop}) should be 
multiplied by four.
The total EDM of the fermion $F$ from the Barr-Zee type diagrams
 with R-parity violating interaction is then
\begin{eqnarray}
d_{F_k} &=& {\rm Im} (\hat \lambda_{ijj} \tilde \lambda^*_{ikk}) \frac{\alpha_{\rm em} n_c Q_f^2 Q_F e}{16\pi^3m_{f_j}} \cdot \left\{ f \left( \tau \right) - g\left( \tau \right) \right\} \nonumber\\
&\approx & {\rm Im} (\hat \lambda_{ijj} \tilde \lambda^*_{ikk}) \frac{\alpha_{\rm em} n_c Q_f^2 Q_F e}{16\pi^3m_{f_j}} \cdot \tau \left( 2+ \ln \tau + \cdots \right)\, , \nonumber\\
\label{eq:Barr-Zee1}
\end{eqnarray}
where $\tau = m_{f_j}^2 / m_{\tilde \nu_i}^2$. 
The flavor index, electric charge and number of color of the fermion $F$ are denoted respectively by $k$, $Q_F e$ and $n_c$ ($n_c=3$ if inner loop fermion is a quark, otherwise $n_c=1$), and $j$ and $Q_f e$ are respectively the flavor index and electric charge of the inner loop fermion $f$.
The second line of the above equation is the approximated expression for small $\tau$.
Note also that the Barr-Zee type diagram gives EDM contribution only to down-type quarks, and the same property holds also for the chromo-EDM (cEDM) seen below.

We see from Eq. (\ref{eq:secloop}) (see also Fig. \ref{fig:barr-zee}) that the chirality structure 
of the scalar exchange between internal loop and external line (RPV vertices with sneutrino exchange) 
has the form $P_L \otimes P_R$ and $P_R \otimes P_L$, which is a consequence of the lepton number conservation of the whole EDM process.
This gives as a result the structure 
\begin{equation}
f(\tau)-g(\tau) \approx \tau ( 2 + \ln \tau ) \ ,
\label{eq:f-g}
\end{equation}
in the final formula (\ref{eq:Barr-Zee1}). 
This is consistent with the result obtained in the analysis 
of the Barr-Zee type diagram analogues with the exchange of Higgs 
bosons in the two Higgs doublet model, done originally by Barr 
and Zee \cite{barr-zee} (see also \cite{barger}).
In the two Higgs doublet model, there are also additional 
contributions with the structures $P_L \otimes P_L$ and 
$P_R \otimes P_R$ which yield contribution proportional to 
\begin{equation}
f(\tau) + g(\tau) \approx \tau \left( \frac{\pi^2}{3} + 2 + \ln \tau + (\ln \tau)^2 \right) \, ,
\end{equation}
which is absent in the RPV supersymmetric models.

The small $\tau$ behaviour in Eq. (\ref{eq:Barr-Zee1}) is in 
contradiction with the result presented in 
Refs. \cite{chang,herczege-n,faessler}, where the RPV Barr-Zee type 
diagrams receive the leading contribution proportional to $\tau (\ln \tau)^2$.
If one would replace $f(\tau)-g(\tau)$ 
in Eq. (\ref{eq:Barr-Zee1}) by $f(\tau)+g(\tau)$, 
the formula given in Refs. \cite{chang,herczege-n,faessler}
would be obtained.
The difference between these two results is large.
For example, 
 if we consider the Barr-Zee type diagram with the 
bottom quark and tau lepton in the loop,
the electron EDM evaluated from our formula
in Eq. (9) is one order of magnitude smaller than those of Refs. 
\cite{chang,herczege-n,faessler}
and even sign of the electron EDM is different as shown in Table I.
By using our correct formula, the experimental upper bounds on 
RPV interactions given from the RPV Barr-Zee type contribution is 
loosened by one order of magnitude.
\begin{table}
\begin{tabular}{ccccc}
\hline
\hline
$m_{\tilde{\nu}}$   &    tau lepton & & bottom quark &   \\ 
{}  [TeV]      &   $f-g$ (ours)  &  $f+g$  &  $f-g$ (ours) & $f+g$ \\ \hline
 0.1 &  $3.14 $              &   $ -32.1 $            & $1.80 $             & $-15.9 $ \\
 1   &  $5.50 \times 10^{-2}$ &   $-7.89\times 10^{-1}$ & $3.64 \times 10^{-2}$ & $-4.64\times 10^{-1} $\\
 5   &  $2.87 \times 10^{-3}$ &   $-4.99\times 10^{-2}$ & $1.98 \times 10^{-3}$ & $-3.12 \times 10^{-2}$\\ \hline
\end{tabular}
\caption{The electron EDM $d_e$ [$10^{-27}$e cm]. 
The coupling constants of RPV interactions are set to
Im$(\lambda_{233}\lambda_{211}^*)= $Im $(\sum_i \lambda_{i33}'\lambda_{i11}^*)= 10^{-5}$
and the masses of b-quark and tau lepton are set to $m_b=4.2$ GeV and $m_\tau=1.78$ GeV.
For comparison, we have shown the EDM calculated by replacing $f-g$ by $f+g$.
}
\end{table}

We can also evaluate Barr-Zee type diagrams which contribute to the quark cEDM. The lagrangian of the cEDM interaction is given by 
\begin{equation}
{\cal L}_{\rm cEDM} = -i  \frac{d^c_q}{2} \bar \psi \gamma_5 \sigma^{\mu \nu} T_a \psi F^a_{\mu \nu } 
\, ,
\end{equation}
where $F_{\mu \nu}^a$ is the gluon field strength.
The Barr-Zee type contribution of the down-type quark $q_k$ is then 
\begin{equation}
d^c_{q_k} = {\rm Im} (\lambda'_{ijj} \lambda'^*_{ikk}) \frac{\alpha_{\rm s} g_s}{32\pi^3m_{q_j}} \cdot \left\{ f \left( \tau \right) - g\left( \tau \right) \right\} \, ,
\label{eq:chromo-Barr-Zee1}
\end{equation}
where $\tau = m_{q_j}^2 / m_{\tilde \nu_i}^2$.
The flavor indices of the quark $q_k$ and the quark of the inner loop are denoted respectively by $k$ and $j$.

In conclusion, we have reanalyzed the RPV supersymmetric contribution to the Barr-Zee type two-loop diagrams, and have found that the result gives smaller fermion EDM, by one order of magnitude than the previous analyses \cite{chang,herczege-n,faessler} (for sneutrino mass = 1 TeV).
This difference is significant, as it can alter the relative size between other contributing processes such as the 4-fermion interactions \cite{herczege-n,faessler}.
Nevertheless our finding does not alter the dominance of the Barr-Zee type diagrams over the other two-loop diagrams, as shown in the analysis of Chang {\it et al.} \cite{chang}, and their conclusion is still very important.

\end{document}